\documentclass[twocolumn,showpacs,preprintnumbers,amsmath,amssymb,a4paper,superscriptaddress]{revtex4}
\usepackage{amsmath}
\usepackage{graphicx}
\usepackage{epstopdf} 
\usepackage{dcolumn}
\usepackage{graphics}      
\usepackage{subfig}  

\begin{document}
\title{Investigation of Confinement Induced Resonance in Atomic Waveguides with Different Geometries by Quantum Monte Carlo Methods}
\date{\today}
\author{Sajad Azizi}
\email[]{sazizi@iasbs.ac.ir; sajjazizi@gmail.com}
\author{Shahpoor Saeidian}
\email[]{saeidian@iasbs.ac.ir}
\affiliation{Department of Physics, Institute for Advanced Studies in Basic Sciences
(IASBS),
Gava Zang,
Zanjan 45137-66731,
Iran}

\begin{abstract}\label{txt:abstract}
We have investigated the quantum dynamics of two ultracold bosons
inside an atomic waveguide for two different confinement geometries (cigar-shaped and toroidal waveguides) by quantum Monte Carlo methods.  For quasi-1D gases, the confining potential of the waveguide leads to the so-called confinement induced resonance (CIR), results in the phase transition of the gas to the impenetrable bosonic regime (known as TG gas). In this regime the bosons repel each other strongly and behave like fermions. We reproduce CIR for a cigar-shaped waveguide and analyze the behavior of the system for different conditions.  Moreover, our analysis demonstrates appearance of CIR for a toroidal waveguide.  Particularly, we show that the resonance position is dependent on the size of the waveguide, which is in contrast to the cigar shaped waveguides for which it is universal.  
 \end{abstract}

\maketitle

\section{Introduction\label{sec:introduction}}
By invention of the intellectual techniques for cooling atoms to nano Kelvin, the ultracold atomic gases have been investigated widely during the last two decades.  In this regime, the thermal wavelength of the atoms 
$\lambda_T$, is less than their average distance and the wave nature affects the statistical behavior of the gas, such that the classical physics is unable to explain observed phenomena.
As an example one can refer to the Bose-Einstein condensate \cite{Pethick02}. Using laser and external fields, nowadays we are able to localize atoms in different types of optical lattices \cite{Bloch08}. On the other hand the Feshbach resonance allows us  to control the atom-atom interaction \cite{Bloch05,Lewenstein}. With these possibilities, ultracold atoms can be used for simulation of solid state systems, quantum computation, quantum metrology and so on. It is also possible to confine the atoms in one or more dimensions, which provides us the exceptional possibilities to study low dimensional quantum systems \cite{Bloch08}.

The quantum dynamics of low-dimensional systems are strongly influenced by the confinement geometry. As an example, confining atoms in two dimensions leads to the so-called confinement induced resonance (CIR) which was first predicted for s-wave bosonic scattering by M. Olshanii in his seminal work \cite{Olshanii98}. It is a kind of zero energy Feshbach resonance, and occurs if the binding energy of the quasi-molecular state of the atom-atom interaction potential coincides with the energy spacing between the levels of the transverse confining potential.  This leads to the resonant molecule formation due to the coupling between the ground transverse mode and the manifold of the excited modes \cite{Bergeman03}.

In quasi-1D atomic gases, in the vicinity of the CIR,  the coupling constant between the atoms diverges, which can result  in a total atom-atom reflection, thereby creating a gas of impenetrable atoms known as TG gas. In this case the strongly interacting bosons behave like weakly interacting fermions such that the atoms can not occupy the same place. 

CIRs were confirmed in several experiments \cite{Kinoshita04,Paredes04,Guenter05,Haller09,Haller10,Lamporesi10,Froehlich11}.  
Theoretical study of CIRs have been done for bosonic s-\cite{Bergeman03,Olshanii2011,Melezhik07,Saeidian08,Saeidian12} and d-wave \cite{Giannakeas11} and dipolar \cite{Sinha07,Giannakeas13} scattering , fermionic p-wave scattering \cite{Melezhik07,Granger04,Saeidian15}, as well as distinguishable atom scattering \cite{Kim06,Kim07,Melezhik09} in single- and multi- (transverse)mode regime\cite{Saeidian08,Melezhik11,Moore04,Shadmehri16}.  In spite of intensive theoretical studies of CIRs, the existing theoretical models need to be improved for a quantitative interpretation and guiding the experiments.  All these models consider an ideal quasi-1D system in which the atoms are confined in the transverse directions while they can move freely in the longitudinal direction.  The main goal of the present work is to consider more realistic confinement.  By employing the quantum Monte Carlo methods DMC and F-NDMC \cite{Anderson, Kosztin, Dugan}, we  study quantum dynamics of two ultracold bosons in a waveguide and analyze the resonance position of the system.  We consider two different geometries for the waveguide: the cigar-shaped and the toroidal geometry.  To the best of our knowledge CIR has not already been studied for toroidal waveguides.

This paper is organized as follows.  In section \ref{sec:Model} we describe the Hamiltonian of our system. The numerical methods DMC and F-NDMC are explained in section \ref{sec:NU}.  We show our results in section \ref{sec:RE}.  Finally we summarize and conclude with section \ref{sec:conclusion}.

\section{Hamiltonian of two atoms in a waveguide\label{sec:Model}}
Let us assume two identical atoms with mass $m$ inside a waveguide described by the confining potential $ V_{C}(\textbf{r})$. The Hamiltonian of this system is given by 

\begin{equation}
H = \sum_{i=1}^{2} \Big(-\dfrac{\hbar^{2}}{2m}\nabla^{2}_{i} + V_{C}(\textbf{r}_i)\Big) + V(\textbf{r}_{1},\textbf{r}_{2}),
\label{e1}
\end{equation}
where $V(\textbf{r}_{1},\textbf{r}_2)$ is the inter-atomic interaction potential which is assumed as follows

\begin{align}
V(\textbf{r}_1,\textbf{r}_2)=
\begin{cases}
\infty &  \text{$|\textbf{r}_1 -\textbf{r}_2 |  \leq r_c$}\\
-\frac{V_0}{|\textbf{r}_1 -\textbf{r}_2 |^6}& \text{$|\textbf{r}_1 -\textbf{r}_2 |> r_c$},
 \end{cases}
\label{e2}
\end{align}
$ r_{c} $ is a constant.

Performing the transformation $r\rightarrow r/\overline{a},  E\rightarrow E/E_{0}, \omega \rightarrow \omega/\omega_{0}$ and $V \rightarrow V/E_{0} $ with $ E_{0} = \hbar^{2}/m\overline{a}^{2} , \omega_{0} = E_{0}/\hbar$ and $ \overline{a} = 4\pi\Gamma(1/4)^{-2}R_{vdW} $($ R_{vdW} $  and $\Gamma(x)$ are the van der Waals radius and gamma function, respectively) we arrive at the rescaled Schr\"odinger equation:

\begin{align}
\Big\{\sum_{i=1}^{2}\Big(-\dfrac{1}{2}\nabla_{i}^{2} + V_{C}(\textbf{r}_i)\Big) + V(\textbf{r}_1,\textbf{r}_2)\Big\}\Psi(\textbf{r}_1,\textbf{r}_2)\nonumber\\ 
 = E\Psi(\textbf{r}_1,\textbf{r}_2).
\label{e3}
\end{align}
We investigate two different kinds of waveguide:
\subsection{Cigar-Shaped Waveguide \label{subsec:CSW}}
\begin{figure}[!htbp]
\includegraphics[scale=0.35]{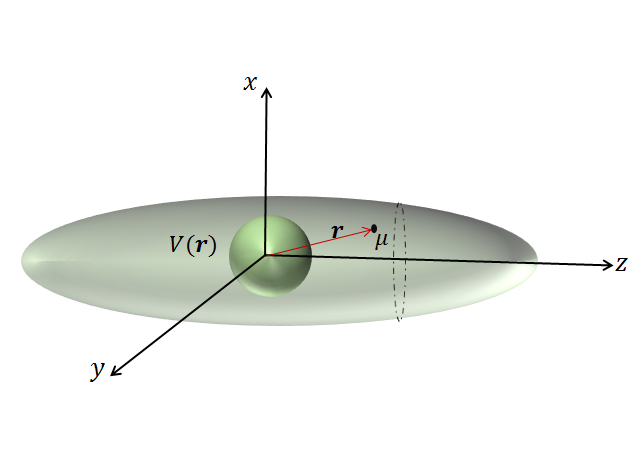}
\caption{\label{fig:f1} Schematic drawing of a cigar-shaped waveguide.}
\end{figure}

A cigar-shaped waveguide (Fig.\ref{fig:f1}) can be approximated by the following potential 
\begin{equation}
V_{C}(\textbf{r}) = \dfrac{1}{2}\omega^{2}(x^{2} + y^{2} + \lambda^{2}z^{2}) \quad \lambda < 1.
\label{e4}
\end{equation}
In the center of mass ($\textbf{R} = \frac{\textbf{r}_1+\textbf{r}_2}{2}$) and relative ($ \textbf{r} = \textbf{r}_1 - \textbf{r}_2$) coordinates, the Hamiltonian can be written as
\begin{equation}
H = H_{CM} + H_{rel},
\label{e5}
\end{equation}
where 
\begin{equation}
\begin{array}{c} H_{CM} = -\dfrac{1}{2M}\nabla_{R}^{2} + \dfrac{1}{2}M\omega^{2}(X^{2}+Y^{2}+\lambda^{2}Z^{2}),\\
H_{rel} = -\dfrac{1}{2\mu}\nabla_{r}^{2} + \dfrac{1}{2}\mu \omega^{2}(x^{2}+y^{2}+\lambda^{2}z^{2})+V(\textbf{r}), \end{array}
\label{e6}
\end{equation}
and $M=2 $ and $ \mu =1/2 $ are the total mass and reduced mass respectively.

The form of the Hamiltonian allows the separation of the CM and relative motions $\Psi(\textbf{r}_1,\textbf{r}_2) = \Psi_{CM}(\textbf{R})\Psi_{rel}(\textbf{r})$, where $ \Psi_{CM}(\textbf{R}) $ and $ \Psi_{rel}(\textbf{r}) $ obeys the following Schr\"odinger equation:

\begin{align}
\Big \{-\dfrac{1}{2M}\nabla_{R}^{2} + \dfrac{1}{2}M\omega^{2}(X^{2}+Y^{2}+\lambda^{2}Z^{2})\Big \}\Psi_{CM}(\textbf{R})\nonumber\\ = E_{CM}\Psi_{CM}(\textbf{R}),\nonumber\\
\Big \{-\dfrac{1}{2\mu}\nabla_{r}^{2} + \dfrac{1}{2}\mu \omega^{2}(x^{2}+y^{2}+\lambda^{2}z^{2})+V(\textbf{r})\Big \}\Psi_{rel}(\textbf{r})\nonumber\\ = E_{rel}\Psi_{rel}(\textbf{r}),
\label{e7}
\end{align}
and $E = E_{CM}+E_{rel}$. The Schr\"odinger equation for the relative motion describes scattering of a particle with mass $\mu$ off the central potential $ V(\textbf{r}) $ under the 3D harmonic potential.

The analytical \cite{Olshanii98,Olshanii2011} and numerical \cite{Saeidian08,Saeidian12} studies for s-wave bosonic scattering of two atoms with mass $m$ under 2D harmonic confinement ($\lambda=0$) with frequency $\omega$,  predict a universal position for the CIR at $a_s/a_{\perp}=1/C=0.68$, corresponding to divergence of the coupling constant between the atoms.  Here $a_s$ is the s-wave scattering length, $a_{\perp} = \sqrt{2\hbar/m\omega}$ is the transverse oscillator length, and $C=-\zeta(1/2)$.  $\zeta(x)$ is the Zeta function.  In this case the atoms repel each other strongly and we have $\Psi_{rel}(\textbf{r}=0)=0$ (i.e the atoms behave like fermions and do not occupy the same place).  In this work we show that the same effect can be achieved for $\lambda\neq 0$.

\subsection{Toroidal Waveguide \label{subsec:DSW}}
\begin{figure}[!htbp]
\includegraphics[scale=0.35]{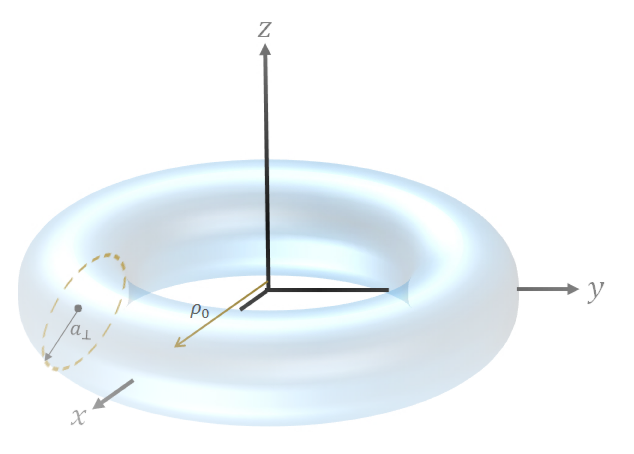}
\caption{\label{fig:f2} Schematic drawing of a toroidal waveguide.}
\end{figure}

The toroidal waveguide can be described by the following confining potential
\begin{align}
V_C(\textbf{r}) = \frac{1}{2}m\omega^2 \Big[(\rho-\rho_0)^2+z^2\Big],
\label{e8}
\end{align}
where $\rho = (x^2+y^2)^{\frac{1}{2}}$ and $\rho_0$ is the radius of the waveguide(Fig.\ref{fig:f2}).

Due to the symmetry of the Hamiltonian, one can write in the cylindrical coordinate $(\textbf{r}=(\rho,\varphi,z))$:
\begin{align}
\Psi(\rho_1,\varphi_1,z_1;\rho_2,\varphi_2,z_2)=\Psi(\rho_1,\varphi_1-\varphi_2,z_1;\rho_2,0,z_2).
\label{e9}
\end{align} 

We show that CIR can occur in toroidal waveguide as well.  However, in this case the resonance position $a_s^{CIR}/a_{\perp}$ is not universal and depends on the size of  the waveguide.

\section{Numerical Method\label{sec:NU}}
In this work we have employed the numerical methods DMC and F-NDMC to solve the Schr\"odinger equation for two atoms in a waveguide.  The DMC is a method to calculate the ground state $\Psi_0(\textbf{R})$ of the Hamiltonian H of a N-particle system by assuming the trial wavefunction $\Psi_T(\textbf{R})$ and considering propagation of the function $f(\textbf{R},\tau) = \Psi_T(\textbf{R})\Psi_0(\textbf{R},\tau)$ in imaginary time $\tau =it/\hbar$. Here $\textbf{R} = \{\textbf{r}_1, ..., \textbf{r}_N \}$ denotes the position of the particles in the configuration space. 
In this method we consider a certain number of randomly distributed walkers in the configuration space
and let the density of the walkers to evolve according to the time dependent Schr\"odinger equation for $f(R,\tau)$.  The density of the walkers tends to $f(R)=\psi_T(R)\psi_0(R)$ after a number of iteration.  By knowing $f(R)$ and $\psi_T(R)$ we can calculate the wavefunction $\psi_0(R)$.

The bosonic function must be symmetric with respect to exchange of the two particles. The most natural
way to construct the trial wavefunction $\Psi_T$ is to consider a product of one-body
and two-body terms:
\begin{align}
\Psi_T(\textbf{r}_1, ..., \textbf{r}_N)= \Pi_{i=1}^{N}f_1(\textbf{r}_i)\Pi_{j<k}^{N}f_2(|\textbf{r}_j-\textbf{r}_k|)
\end{align}
This construction is called the Bijl-Jastrow trial wavefunction.  Here the one-body term $f_1(\textbf{r})$ accounts
for the external potential.  It is the solution of the Sch\"odinger equation for a single particle in the same external potential. The interaction between the particles is accounted by the two-body Bijl-Jastrow term $f_2(r)$.  

The DMC method can not be used directly to calculate the excited states of the system. The reason is that in this case the function $f(\textbf{R},\tau)$ is not positive everywhere and can not be interpreted as a probability density. The F-NDMC algorithm is a modified DMC method which is able to calculate the excited state $\Psi$ of the system, by enforcing the positive definiteness of the function $f(\textbf{R},\tau)$.

\section{Results\label{sec:RE}}
Using DMC and F-NDMC methods we have solved the Schr\"odinger equation for two bosons in an atomic waveguide and found the eigenfunctions and eigenenergies of the system for different conditions.

\subsection{Cigar-Shaped Waveguide\label{subsec:RCSW}}
The corresponding Schr\"odinger equation in this case is given by 
\begin{align}
\Bigg \{ \sum_{i =1}^2 \Big(-\frac{1}{2m}\nabla_i^2 +\frac{1}{2}m\omega^2(x_i^2 + y_i^2 + \lambda z_i^2)
\Big) \nonumber\\  +V(\textbf{r}_1,\textbf{r}_2)
 \Bigg \}
 \Psi(\textbf{r}_1,\textbf{r}_2)
  = E\Psi(\textbf{r}_1,\textbf{r}_2).
\label{e11}
\end{align}
 We assume the Bijl-Jastrow trial wavefunction as follows:
\begin{align}
\Psi_T(\textbf{R})= (z_2+z_1)\Pi^2_{i=1} e^{-\frac{1}{2}\omega(x^2_i+y^2_i+\lambda z^2_i)}\nonumber\\ \times \Big(J_{1/4}(\frac{r_0^2}{2r^2})+\alpha J_{-1/4}(\frac{r_0^2}{2r^2})\Big),
\label{e12}
\end{align}
where $\alpha =-J_{1/4}(r_0^2/2r_c^2)/J_{-1/4}(r_0^2/2r_c^2)$, $r=|\textbf{r}_1-\textbf{r}_2|$ and $r_0=(2\mu V_0)^{1/4}$.  The one body terms describe two atoms in the ground state of the harmonic potential in the transverse directions.  Besides, the center of mass is in the first excited state of the harmonic potential in the longitudinal direction.

\begin{figure}[!htbp]
\includegraphics[scale=0.3]{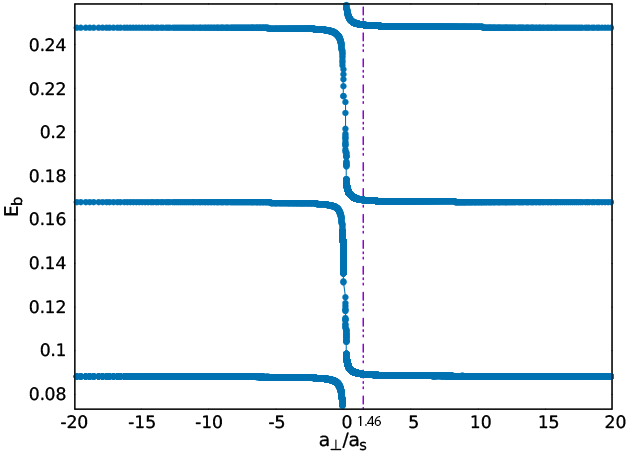}
\caption{\label{fig:f3} The binding energy $E_b$ as a function of $a_{\perp}/a_s$ for $\omega=0.04$ and $\lambda=0.2$.}
\end{figure}

Fig.\ref{fig:f3} shows our results for the binding energy $E_b$ of the quasi-molecular state as a function of $a_{\perp}/a_s$ for up to three open channels for $\omega_{\perp}=0.04$ and $\lambda = 0.2$.  There is a bound state for each transverse channel.    While in free space the molecular state exists just for positive values of the scattering length $a_s$, in a waveguide such a state exists for all $a_s$.   This is in agreement with analytical results \cite{Bergeman03}.  The transverse channel thresholds in the figure correspond  to $V_0=0$ (no interaction) and $L_z=0$, and are given by
\begin{align}
E_n = 2(n+1+\frac{\lambda}{2})\omega,
\label{e13}
\end{align}
where $n = 0,1,2,...$ labels the transverse channels.  With increasing $V_0$, the binding energy $E_b$ decreases abruptly when $a_s$ diverges ($a_{\perp/a_s}\rightarrow 0$). 
CIR occurs when the binding energy $E_b$ of an excited channel coincides with the energy threshold of the lower channel.

\begin{figure}[!htbp]
\includegraphics[scale=0.45]{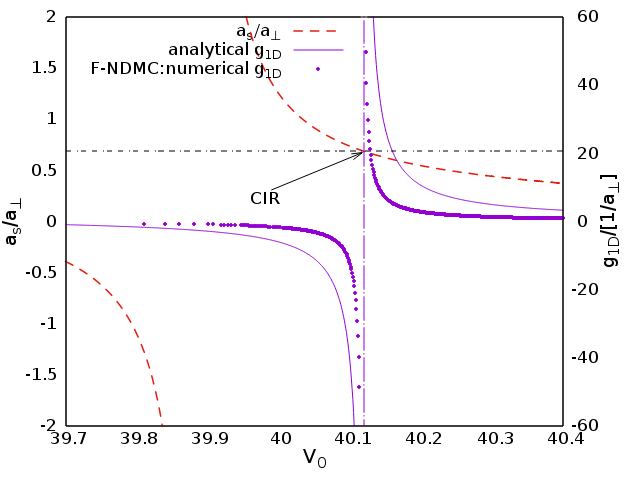}
\caption{\label{fig:f4} The coupling constant and scattering length as a function of the potential depth $V_0$ for $\omega=0.04$ and $\lambda=0.2$.}
\end{figure}

We can eliminate the transverse degrees of freedom and approximate the Schr\"odinger equation (\ref{e7}) for the relative coordinate to an effective 1D equation with $z$ the only variable:
\begin{align}
\Big \{-\dfrac{1}{2\mu}\dfrac{d^2}{dz^2} + \dfrac{1}{2}\mu \lambda^{2}\omega^{2}z^{2}+g_{1D}\delta(z)+\omega\Big \}\Psi_{1D}(z)\nonumber\\ = E\Psi_{1D}(z),
\label{1D equation}
\end{align}
with $E=E_{rel}$, provided that the 1D coupling constant $g_{1D}$ be chosen as
\begin{align}
g_{1D}=\frac{1}{2\mu}\frac{\Psi_{rel}'(0^{+})-\Psi_{rel}'(0^{-})}{\Psi_{rel}(0)},
\label{g1D_NUM}
\end{align}
where $\Psi_{rel}'(0^{\pm})=\frac{\partial}{\partial z}\Psi_{rel}(\textbf{r})\big |_{x=y=0, z=0^{\pm}}$.
On other hand in \cite{Olshanii98} it has been shown (for $\lambda=0$) that $g_{1D}$ must obey the following relation 
\begin{align}
g_{1D}=\frac{2}{\mu a_{\perp}}\Big(1-C\frac{a_s}{a_{\perp}}\Big)^{-1}.
\label{CouplAnal}
\end{align}
In Fig.\ref{fig:f4} we have depicted our results for $g_{1D}$  as a function of the potential depth $V_0$ along with the analytical result (\ref{CouplAnal}) for comparison.  The rescaled scattering length $a_s(V_0)/a_{\perp}$ is also provided.  It is obvious that the coupling constant $g_{1D}$ diverges at $V_0\approx 40.12$ for which $a_s/a_{\perp}=1/C=0.68$.  The numerical results are in relatively good agreement with the analytical prediction (\ref{CouplAnal}).

\begin{figure}[!htbp]
\subfloat[Subfigure 1 list of figures text][]
        {
\includegraphics[width=0.25\textwidth]{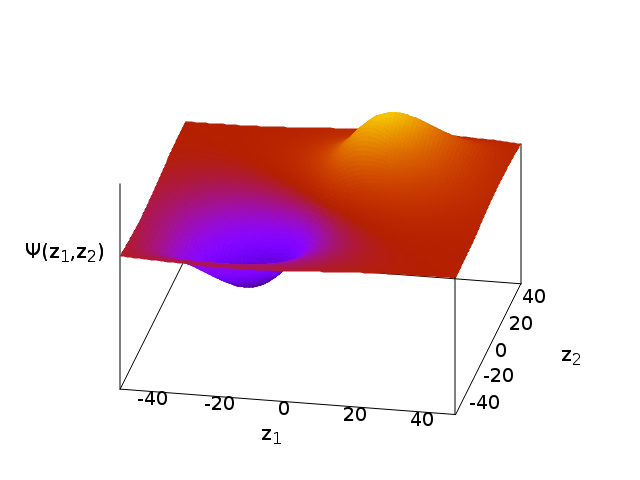}
\label{fig:subfig1}
        }
\subfloat[Subfigure 2 list of figures text][]
        {
\includegraphics[width=0.25\textwidth]{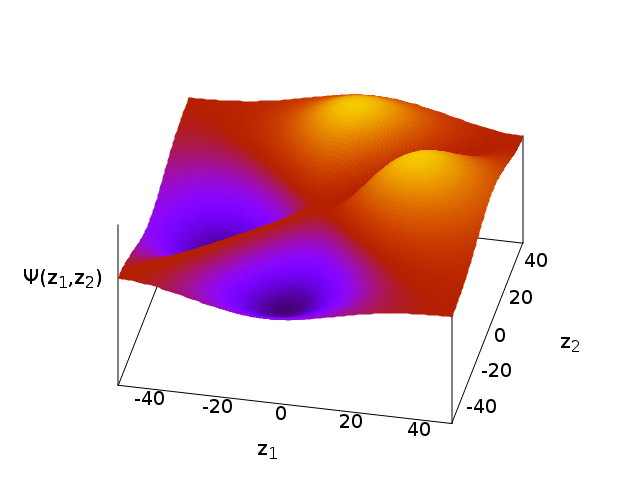}
\label{fig:subfig2}
        }
        \\
            \subfloat[Subfigure 3 list of figures text][]
        {
        \includegraphics[width=0.25\textwidth]{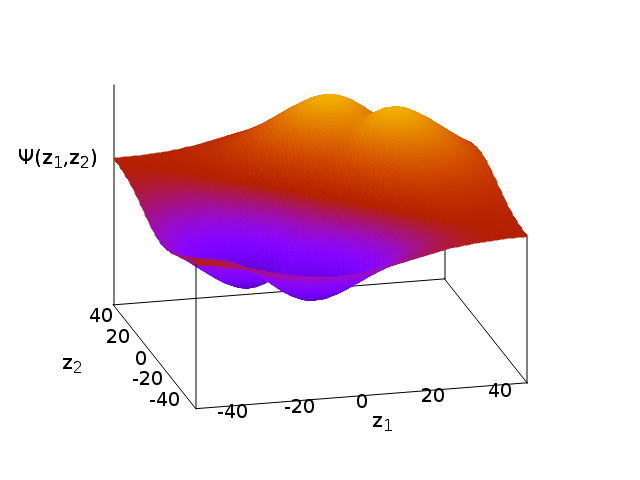}
        \label{fig:subfig3}
        }
            \subfloat[Subfigure 4 list of figures text][]
        {
        \includegraphics[width=0.25\textwidth]{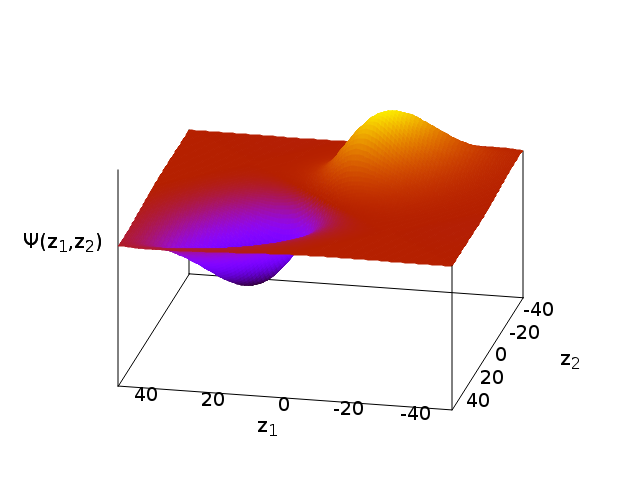}
        \label{fig:subfig4}
        }
    \caption{The two-particle wavefunction with respect to the atoms position $z_1$ and $z_2$ for $x_1=x_2=y_1=y_2=0$ for
 $a_s/a_{\perp} = 0.3083$ (a),  
 $a_s/a_{\perp} = 0.6857$ (b),
 $a_s/a_{\perp} = 0.9608$ (c), and
 $a_s/a_{\perp} = 3.5479$ (d).
    }
\label{fig:f5}
\end{figure}
Fig.\ref{fig:f5} shows the two-particle wavefunction $\Psi(\textbf{r}_1,\textbf{r}_2)$ in $z_1-z_2$ plane ($x_1=x_2=y_1=y_2=0$) for different values of $a_s/a_{\perp}$. As we expect, $\Psi$ is symmetric with respect to interchange particles $1$ and $2$. Far from the CIR, $g_{1D}$ is negligible. 
In this case the wavefunction includes a maximum and a minimum (Figs. \ref{fig:f5}-(a) and -(d)).   However, near the CIR each of them splits into two parts which is a witness of the strong repulsive force between the atoms (Fig.\ref{fig:f5}-(b) and -(c)).

\begin{figure}[!htbp]
\includegraphics[scale=0.45]{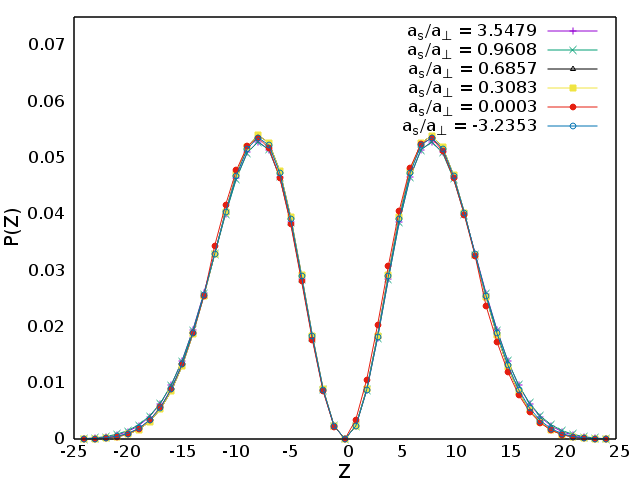}
\caption{\label{fig:f6} The probability distribution for the center of mass with respect to $Z=(z_1+z_2)/2$  (for $X=Y=0$) for different values of $a_s/a_{\perp}$.  These results have been obtained for $\omega=0.04$ and $\lambda=0.2$.}
\end{figure}

\begin{figure}[!htbp]
\includegraphics[scale=0.45]{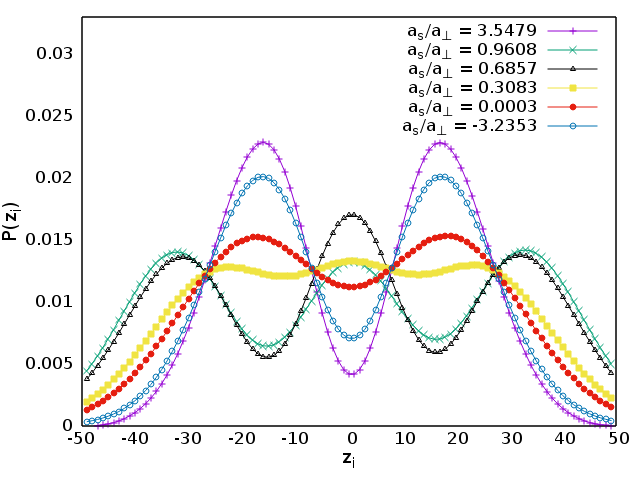}
\caption{\label{fig:f7} The probability distribution vs. atom position $z_i$  (for $x_i=y_i=0$) for $\omega=0.04$ and $\lambda=0.2$  for different values of $a_s/a_{\perp}$.  The position of the another atom has been integrated out.}
\end{figure}

In Fig.\ref{fig:f6} we have shown the probability distribution (along the z-axis) for the CM motion for different values of $a_s/a_{\perp}$.  We can see that the probability distribution is the same for all values of $a_s/a_{\perp}$, which is expectable, because the CM motion is independent of the interaction potential $V(\textbf{r})$.
In contrast, the probability distribution for the relative position $z$ and the atom position $z_i$ is influenced by the interaction potential. 

Fig.\ref{fig:f7} shows the probability distribution $P(z_i)$ vs. the atom position $z_i$. For $a_s=0$ (i.e., no interaction) we see two symmetric peaks.  In this case the atom motion along the $z$ direction is in a superposition of ground and first excited states of the harmonic potential ($|n_z^1=0\rangle |n_z^2=1\rangle+|n_z^1=1\rangle |n_z^2=0\rangle$).  By increasing $a_s$, a new peak appears at $z=0$.  Besides, the distance between the outer peaks increases.  At the CIR the peak at $z=0$ has the maximum amplitude and the distance between the outer peaks becomes maximum (i.e. the gas is expanded).  By increasing $a_s$ further, the inner peak loses amplitude and finally disappears and $P(z_i=0)$ tends to zero, while the outer peaks become closer to each other and increase in height.  At the limit $|a_s|\rightarrow \infty$ we see two narrow peaks due to strong attractive force between the atoms which leads to formation of a quasi-molecular state.  This is in agreement with results of \cite{Saeidian12}.

\begin{figure}[!htbp]
\includegraphics[scale=0.45]{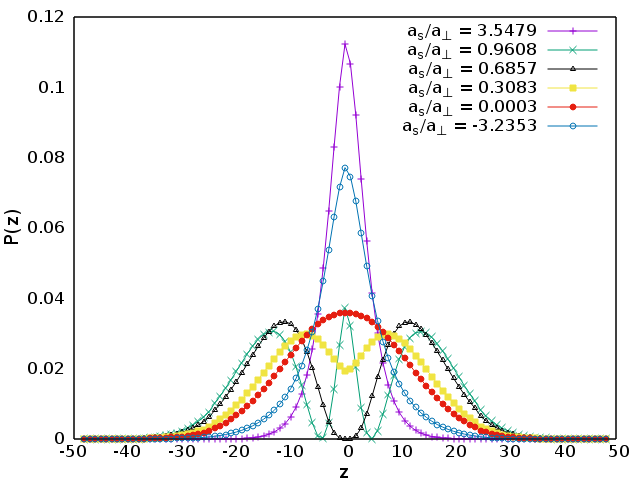}
\caption{\label{fig:f8} The probability distribution for the relative motion vs. $z=z_2-z_1$ (for $x=y=0$) for different values of $a_s/a_{\perp}$.  These results have been obtained for $\omega=0.04$ and $\lambda=0.2$.}
\end{figure}

In Fig.\ref{fig:f8} we have depicted the probability distribution $P(z)$ vs. $z$ for the relative motion.  In the case of $a_s\approx 0$ (i.e. no interaction) the reduced mass particle is in the ground state of the harmonic potential and the probability distribution $P(z)$ is Gaussian.  By increasing the scattering length, the peak splits into two parts.  In the vicinity of the CIR ($a_s/a_{\perp}\sim 1$), the coupling constant between the atoms diverges and they repel each other strongly.  In this case we see two well separated peaks in the plot of $P(z)$.  By increasing $a_s/a_{\perp}$ further, these two peaks lose amplitude and then disappear and a narrow peak appears at $z=0$, which corresponds to strong attractive force between the atoms.  For $a_s<0$ as it is expectable we see just one peak which increases in hight and becomes narrower as $a_s\rightarrow -\infty$.

\subsection{Toroidal Waveguide\label{subsec:RDSW}}
\begin{figure}[!htbp]
\includegraphics[scale=0.30]{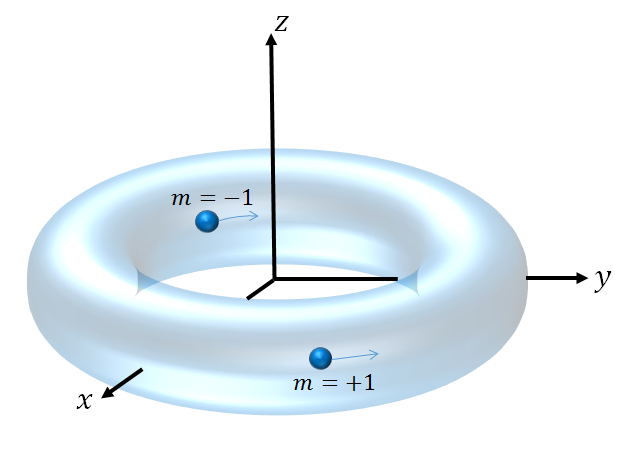}
\caption{\label{fig:f9} The schematic drawing of two particles in a toroidal waveguide.}
\end{figure}

The Schr\"odinger equation for this waveguide is given by 
\begin{align}
\Bigg \{ \sum_{i =1}^2 \Big(-\frac{1}{2m}\nabla_i^2 +\frac{1}{2}m\omega^2((\rho_i - \rho_0)^2 + z_i^2)
\Big) \nonumber\\  +V(\textbf{r}_1,\textbf{r}_2)
 \Bigg \}
 \Psi(\textbf{r}_1,\textbf{r}_2)
  = E\Psi(\textbf{r}_1,\textbf{r}_2).
\label{e17}
\end{align}
For our calculations we use the following trial wavefunction
\begin{align}
\psi(\textbf{r}) = \cos [ m(\varphi_2 - \varphi_1)]e^{-\frac{1}{2} \omega\sum_{i=1}^2 [(\rho_i - \rho_0)^2 + z_i^2 ]}\nonumber\\ \times \Big(J_{1/4}(\frac{r_0^2}{2r^2})+\alpha J_{-1/4}(\frac{r_0^2}{2r^2})\Big),
\label{e18}
\end{align}
which is symmetric under the interchange of the two particles. 
This wave function describes two particles in the ground state of the 2ِD harmonic potential ( in the $\rho$-$z$ plane), which are rotating in opposite directions around the $z$-axis, with angular momentum $L_z=m$ (Fig.\ref{fig:f9}).  We assume $|m|=1$.  

\begin{figure}[!htbp]
    \centering
       \subfloat[subfigure 1 list of figures text][]
        {
        \includegraphics[width=0.35\textwidth]{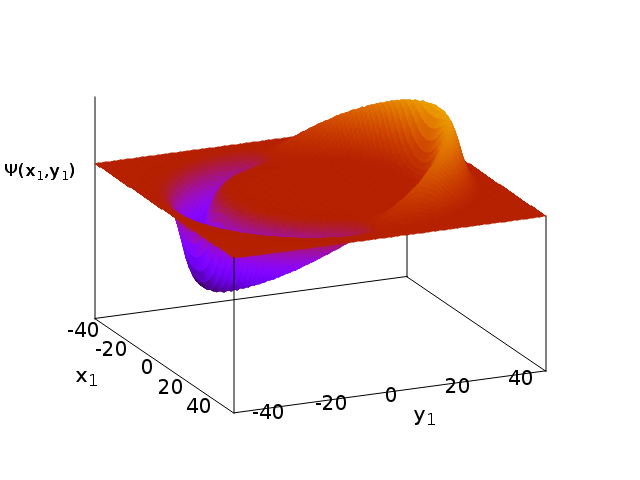}
        \label{fig:subfig1}
        }

       \subfloat[subfigure 2 list of figures text][]
        {
        \includegraphics[width=0.35\textwidth]{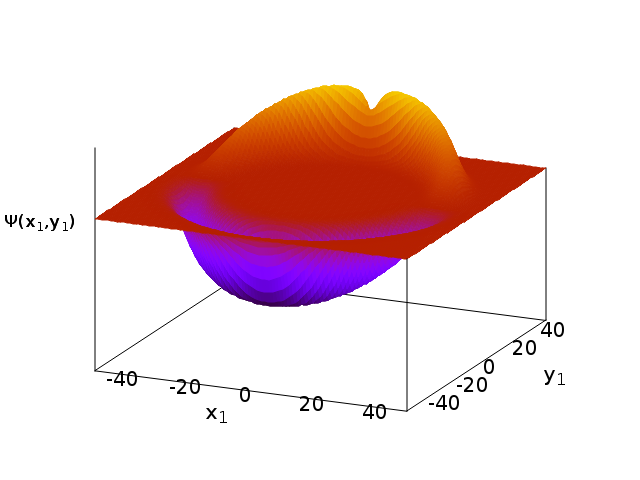}
        \label{fig:subfig2}
        }
    \caption{The two-particle wavefunction $\Psi(\textbf{r}_1,\textbf{r}_2)$ in $x_1-y_1$ plane for $x_2=\rho_0+\delta$ and $y_2=z_2=0$ with $\omega=0.01$, $\rho_0=112$ and $\delta=5r_c$ for  $V_0=0$ (a) and  $V_0=40.65$  (b). 
    }
\label{fig:f10}
\end{figure}

\begin{figure}[!htbp]
\includegraphics[scale=0.45]{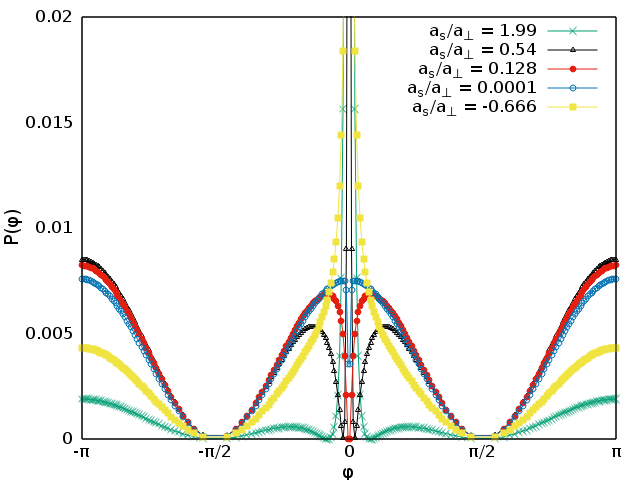}
\caption{\label{fig:f11} The probability distribution $P$ as a function of $\varphi=\varphi_1-\varphi_2$ for different values of $a_s/a_{\perp}$ for $\omega=0.01$ and $\rho_0=112$.}
\end{figure}

Fig.\ref{fig:f10} shows the two-particle wavefunction $\Psi(\textbf{r}_1,\textbf{r}_2)$ in $x_1-y_1$ plane for $x_2=\rho_0 + \delta$ and $y_2=z_2=0$ for two cases (a) $V_0=0$ (i.e. no interaction) and (b) $V_0=40.65$ (i.e. near to CIR). There are two zeros at $\varphi_1=+\pi/2$ and $-\pi/2$ which is because of the interference between clockwise and counterclockwise rotating wavefunctions. For $V_0=40.65$ the peak at $\varphi_1=0$ splits into two parts which is a witness of the resonance induced by the confinement. 
\\
In Fig.\ref{fig:f11} we have plotted the probability distribution $P(\varphi=\varphi_1-\varphi_2)=|\psi(\textbf{r}_1,\textbf{r}_2)|^2$ for different values of $V_0$ for $\rho_0=112$ and $\omega=0.01$.  Here $\textbf{r}_1$ and $\textbf{r}_2$ have been chosen in cylindrical coordinates $(\textbf{r}=(\rho,\varphi,z))$ as $(\rho_0,\varphi_1,0)$ and $(\rho_0+\delta,\varphi_2,0)$ with $\delta=5r_c$ . 
We observe three different behaviors.  For $|a_s/a_{\perp}|\approx 0$ (i.e. weak interaction) there are wide peaks at $\varphi=0$ and $\varphi=\pm\pi$ (see Fig.\ref{fig:f10}(a)).  For $a_s/a_{\perp}=0.128$, the peak at $\varphi=0$ splits into two peaks which is due to the CIR.  With increasing $|a_s/a_{\perp}|$ a narrow and tall peak appears  at $\varphi=0$ while the peaks at $\varphi=\pm\pi$ loses amplitude and finally disappear.  This corresponds to formation of the quasi-molecular state.

Similar to the cigar-shaped waveguide, it is possible to eliminate the inert degrees of freedom (in this case $\rho$ and $z$) and reduce the problem into a 1D one with $\varphi$ the only variable:
\begin{align}
\Big \{-\dfrac{1}{2\mu\rho_0^2}\dfrac{d^2}{d\varphi^2} + g_{1D}\delta(\varphi)+\omega\Big \}\Psi_{1D}(\varphi)= E\Psi_{1D}(\varphi),
\label{1D equation}
\end{align}
provided that the 1D coupling constant $g_{1D}$ be defined as
\begin{align}
g_{1D}=\frac{1}{2\mu\rho_0^2}\frac{\Psi'(0^{+})-\Psi'(0^{-})}{\Psi(0)},
\label{g1D_NUM}
\end{align}
where $\Psi'(0^{\pm})=\frac{\partial}{\partial \varphi}\Psi(\textbf{r}_1,\textbf{r}_2)\big |_{\varphi_1=\varphi_2\pm 0^{\pm}}$ and $\Psi(0)=\Psi(\textbf{r}_1,\textbf{r}_2)\big |_{\varphi_1=\varphi_2}$.

\begin{figure}[!htbp]
\includegraphics[scale=0.40]{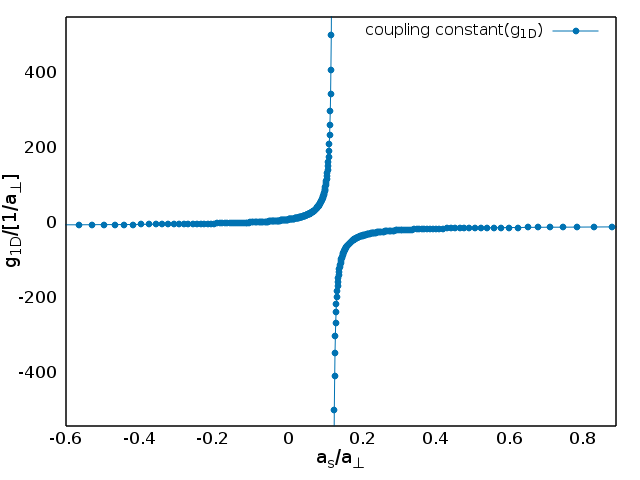}
\caption{\label{fig:f12} The coupling constant $g_{1D}$ as a function of $a_s/a_{\perp}$ for $\rho_0=112$ and $\omega=0.01$.}
\end{figure}

Fig.\ref{fig:f12} shows the coupling constant $g_{1D}$ as a function of $a_s/a_{\perp}$, for $\rho_0=112$ and $\omega=0.01$. It diverges at $a_s/a_{\perp}=0.128$, which corresponds to CIR.

\begin{figure}[!htbp]
\includegraphics[scale=0.40]{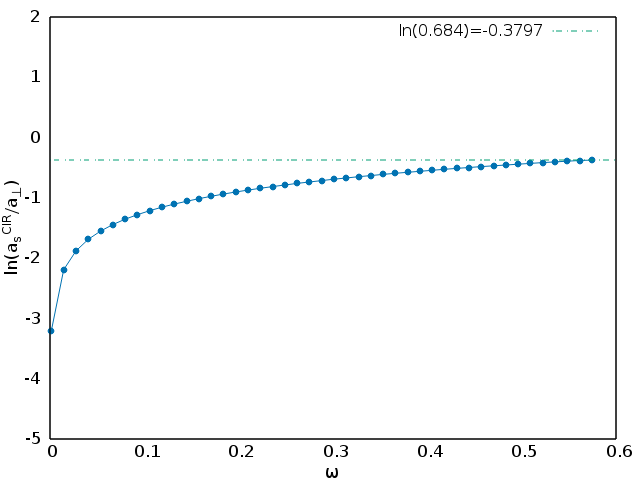}
\caption{\label{fig:f13} The CIR position $a^{CIR}_s/a_{\perp}$ as a function of $\omega$ for $\rho_0=34$.}
\end{figure}

In contrast to the the cigar shaped waveguide for which the resonance position $a_s^{CIR}/a_{\perp}$ is universal ($=1/C=0.68$), for toroidal waveguide it is size dependent.   However as $a_{\perp}$ decreases ($\omega$ increases) it tends to the value $0.68$ (see Fig.\ref{fig:f13} which shows the resonance position as a function of  $\omega$ for $\rho_0=34$).  The larger the radius $\rho_0$, it goes faster to this limit.

\section{Summary and Conclusion\label{sec:conclusion}}
In summary we have studied the quantum dynamics of two ultracold bosons in two different kinds of waveguide (cigar-shaped and toroidal) by quantum  Monte Carlo method. For cigar-shaped waveguide, we showed that the CIR occurs at $a_s/a_{\perp}=0.68$ which has already been obtained for ideal quasi-1D waveguide.  Our calculations demonstrated the occurrence  of CIR for the toroidal waveguide as well.  The CIR occurs at a different position which is dependent on the size of the waveguide.  However, its position $a_s^{CIR}/a_{\perp}$ tends to $0.68$ when $a_{\perp}$ decreases.

\end{document}